\documentclass[letterpaper]{article}
\usepackage{aaai}
\usepackage{courier}
\usepackage{comment}
\usepackage{graphicx}
\usepackage{helvet}
\usepackage{makecell} 
\usepackage{subfig}
\usepackage{times}
\usepackage{xurl}
\usepackage{xcolor, soul}

\newcommand*\samethanks[1][\value{footnote}]{\footnotemark[#1]}

\newcommand{\citet}[1]{\citeauthor{#1} \shortcite{#1}}

\usepackage[hang,flushmargin]{footmisc} 
\begin{document}
%
\title{Manipulating Twitter Through Deletions}

\author{Christopher Torres-Lugo\thanks{Equal contributions.}, Manita Pote\samethanks, Alexander Nwala, and Filippo Menczer\\Observatory on Social Media, Indiana University, Bloomington, USA}

\maketitle

\begin{abstract}
Research into influence campaigns on Twitter has mostly relied on identifying malicious activities from tweets obtained via public APIs. 
These APIs provide access to public tweets that have not been deleted.
However, bad actors can delete content strategically to manipulate the system. 
Unfortunately, estimates based on publicly available Twitter data underestimate the true deletion volume. 
Here, we provide the first exhaustive, large-scale analysis of anomalous deletion patterns involving more than a billion deletions by over 11 million accounts. 
We find that a small fraction of accounts delete a large number of tweets daily.  
We also uncover two abusive behaviors that exploit deletions. 
First, limits on tweet volume are circumvented, allowing certain accounts to flood the network with over 26 thousand daily tweets. 
Second, coordinated networks of accounts engage in repetitive likes and unlikes of content that is eventually deleted, which can manipulate ranking algorithms. 
These kinds of abuse can be exploited to amplify content and inflate popularity, while evading detection.
Our study provides platforms and researchers with new methods for identifying social media abuse.
\end{abstract}

\section{INTRODUCTION}

Originally envisioned as a way to keep in touch with friends and family, social media platforms have  become the primary means for spreading disinformation and conspiracy theories~\cite{Lazer-fake-news-2018}. 
The popularity of social media platforms like Twitter has attracted bad actors seeking to manipulate the attention of millions of users through influence operations. 
These are well documented~\cite{prier2017commanding} and take multiple forms, such as the use of social bots~\cite{ferrara2016rise} to manipulate public opinion~\cite{Shao18hoaxybots} thereby endangering the democratic process~\cite{woolley2018computational}, the economy~\cite{fisher2013syrian}, or public health~\cite{tasnim2020impact,allington2020health,covaxxy-misinfo,infodemic-twitter-vs-facebook}. 
It is therefore more important than ever to study and contain abusive behavior on social media platforms.

Researchers have responded by developing algorithms for identifying different kinds of suspicious activities, for example, accounts controlled by bots \cite{Yang2019botometer,yang2020scalable,botometerv4-2020} and groups of accounts coordinated to accomplish a task \cite{Pacheco2021Coordinated}. 
Most of these research efforts suffer from a common limitation, namely, the reliance on data provided by public APIs that exclude deleted content. 
Bad actors can strategically create and delete high volumes of content to manipulate a platform; the traces of these behaviors may be gone before they can be detected or studied. 
To compound the challenges of detection, platform-wide deletion data is not widely available.

Once a tweet is posted, it could be deleted for multiple legitimate reasons, for example to correct mistakes, protect privacy~\cite{ringel2020proactive}, or in regret~\cite{zhou2016tweet}.
Deletions can be performed by users through the Twitter app, website, third-party services \cite{tweetDelete,tweetDeleteAPI}, and the Twitter API.\footnote{\scriptsize\url{developer.twitter.com/en/docs/twitter-api/v1/tweets/post-and-engage/api-reference/post-statuses-destroy-id}}
In addition, Twitter continuously deletes tweets and accounts that violate platform policies\footnote{\scriptsize\url{help.twitter.com/en/rules-and-policies/enforcement-options}}; all tweets posted by deleted accounts are automatically removed. 

Despite Twitter's enforcement efforts, many actors continue to abuse the platform and evade detection by the strategic use of deletions~\cite{elmas2019power}. 
In this study we present the first systematic, large-scale, and \textit{exhaustive} analysis of anomalous deletion patterns on Twitter. 
This is made possible by leveraging the Twitter Compliance Firehose (CF), a stream of notifications sent to developers that store Twitter data. 
The CF is available only to paying subscribers of the Twitter Enterprise APIs. 
It is meant to be used to honor the expectations and intent of end users.\footnote{\scriptsize\url{developer.twitter.com/en/docs/twitter-api/enterprise/compliance-firehose-api/overview}}
After a tweet or like is deleted, a CF notice is sent to the stream to ensure that developers that stored such content delete their copy within 24 hours. 

We analyzed metadata of about 1.2B deletions by 11.6M accounts, collected from the compliance stream during 30 consecutive days between April 26 and May 25, 2021.  
As detailed in later sections, our research focuses on abusive behaviors involving deletions rather than legitimate uses of deletions. 
Furthermore, to respect user privacy, we consider only deletion metadata and not deleted content. 

This paper makes four primary research contributions:

\begin{enumerate}

    \item The CF service is not freely available, therefore quantitative studies of deletions have relied on estimates based on changes in tweet counts~\cite{infodemic-twitter-vs-facebook,torres2020manufacture}. We measure the extent to which such methods underestimate the true number of deleted tweets.

    \item We provide the first in-depth analysis of compliance data to statistically characterize anomalous deletion behaviors. This is useful in distinguishing between normal deletion patterns and suspiciously high numbers and frequencies of deletions, which could indicate abuse. 

    \item We uncover multiple cases of ongoing abuse on Twitter through the strategic use of deletions. This includes accounts that circumvent Twitter's 2,400 daily tweet limit\footnote{\scriptsize\url{help.twitter.com/en/rules-and-policies/twitter-limits}} through large-volume deletions and coordinated networks of accounts that engage in repetitive likes and unlikes of content that is eventually deleted, which can manipulate ranking algorithms. 

    \item Finally, we characterize accounts engaged in abusive deletion behaviors. We find that frequent and suspicious  deleters tend to have higher bot scores,  profile descriptions related to restricted activities, and mixed suspension rates. 
    
\end{enumerate}

 These findings demonstrate that deletions must receive greater attention by researchers and platforms as a vehicle for online manipulation. 

\section{RELATED WORK}

We can categorize different kinds of social media manipulation into three classes based on whether they are carried out through inauthentic and/or coordinated behaviors, metric inflation, or posting and deleting content. Here, we survey research into influence operations, which has thus far focused on the first two classes. This paper focuses on the last class.

\subsection{Inauthentic and coordinated behaviors}

The use of software agents --- known as social bots \cite{ferrara2016rise} --- to produce content and interact with humans on social media has been studied extensively. 
Early studies of abuse on social media identify spamming as the main goal of early bots.
\citet{yardi2010detecting} demonstrated the existence of spam on Twitter, and found that spam and non-spam accounts differ across network and temporal dimensions. 
Early spambots were simple, having little or no personal information \cite{cresci2020decade}, and easy to detect with strategies that target naive behaviors
\cite{lee2011seven}. 

More recently, as platforms have moved to curtail inauthentic activities, bot operators have responded by evolving their tactics to avoid detection and suspension~\cite{botornot_icwsm17,Yang2019botometer}.
This has led to bots that mimic human-like accounts with detailed personal information and social connections \cite{cresci2020decade}, working as individuals or coordinated groups to influence public opinion or promote discord \cite{broniatowski2018weaponized}.
\citet{boshmaf2011socialbot} demonstrated that social networks like Facebook are vulnerable to large-scale infiltration by social bots. 
Coordination between inauthentic accounts is an effective strategy of amplifying their effect~\cite{zhang2016rise,Pacheco2021Coordinated}. Coordination has been successfully deployed in spreading propaganda and disinformation globally \cite{woolley2016automating}.

\subsection{Metric inflation}

Social media ranking algorithms are tuned to favor popular items because popularity and engagement are widely used as a proxy for quality or authority --- even though this strategy can amplify low-quality content~\cite{ciampaglia2018algorithmic}.
This incentivizes the manipulation of popularity/engagement indicators, such as the number of likes, shares, and followers, as a way to increase exposure, influence, and financial gain. 
This has resulted in a marketplace where the services of accounts that inflate the popularity of a post or individual can be purchased. On Facebook, for example, \citet{de2014paying} studied \textit{like farms}, where likes can be purchased for Facebook pages, revealing that such farms differ in the level of sophistication they employ to evade detection.

\citet{Truthy_icwsm2011class} investigated political astroturf campaigns, whereby an entity controls a group of accounts to create the appearance of widespread support for an individual or cause.
They provided a machine-learning framework leveraging network features to distinguish astroturf campaigns from true political dialogue and revealed multiple instances of such campaigns preceding the 2010 US midterm elections.
\citet{cresci2015fame} proposed a machine-learning method for detecting fake Twitter followers, purchased to create a facade of popularity.
Manipulation is not restricted to automated accounts; \citet{dutta2018retweet} studied the activities of ordinary users who collude in retweeting to boost the popularity of each other's posts. 

\subsection{Posting and deleting}

Research on platform manipulation has primarily used public APIs, which ignore deletions by design --- in order to protect privacy and respect user intent. There are however a few studies that highlight a new form of manipulation in which malicious actors post and delete content to evade detection, which is the focus of this paper.
\citet{elmas2019power} identified the existence of coordinated accounts that publish large volumes of tweets to create a \textit{trending topic}, at which time they delete their tweets to hide the origin of the trend. 
\textit{Follow-train} accounts, which seek to create echo chambers and amplify spam or partisan content, have also been found to publish and delete large volumes of tweets~\cite{torres2020manufacture}.

Malicious deletion behavior is not limited to Twitter; the news media have reported that QAnon-linked YouTube channels delete their videos to avoid content moderation.\footnote{\scriptsize\url{www.cnet.com/features/qanon-channels-are-deleting-their-own-youtube-videos-to-evade-punishment/}}

\begin{table*}
\centering
\caption{Sizes and and descriptions of account categories in the CF stream dataset. We only consider accounts that deleted at least 10 tweets per day.}
\begin{tabular}{lrrrl}
\hline
\textbf{Category}   & \textbf{\% Deletions} & \textbf{\# Accounts} & \textbf{\% Accounts} & \textbf{Description}                  \\ \hline
All deleters        & 100\% & 11,648,492    & 100\% & All accounts in our dataset                                  \\ 
One-day deleters    & 34\% & 7,938,077     &  68\% & Accounts that delete tweets on one day only                      \\ 
30-days deleters    & 4\% & 27,065        & 0.23\%& Account that delete tweets each day in 30-day dataset   \\ 
Suspicious deleters & 1\% & 1,715         &0.015\%&  Accounts that post over the 2,400 daily tweet limit  \\ \hline
\end{tabular}
\label{tab:user_groups_classification}
\end{table*}

\section{DATA COLLECTION}
\label{sec:data_collection}

The Compliance Firehose stream is a service available to subscribers of Twitter's  premium streaming services. 
The stream transmits notifications for state changes to accounts, tweets, and favorites.\footnote{\scriptsize\url{developer.twitter.com/en/docs/twitter-api/enterprise/compliance-firehose-api/guides/compliance-data-objects}} 
Our study leverages the limited metadata included in CF notifications. 
This includes the action  (e.g., tweet deletion or like deletion), the \texttt{id} of the account taking the action, the \texttt{id} of the object of the action, and a timestamp. 
We disregard other user compliance notifications, such as making an account private, removing geotags, or suppressing content in certain countries.

We analyzed the compliance stream for 30 consecutive days, between April 26 and May 25, 2021. 
We honored user privacy by only examining deletion metadata and not deleted content. 
This ensures that our use of the compliance data is consistent with Twitter guidelines. 
Our research aims to focus on abusive behaviors involving deletions rather than legitimate uses of deletions (as discussed in the Introduction). 
To this end, and assuming that the vast majority of small-volume deletions are legitimate, we only consider accounts that delete over 10 tweets per day. This is an admittedly arbitrary threshold, but it does not imply that anyone who deletes more than 10 tweets daily is suspicious; it only excludes from the analysis those who delete less, assuming they are legitimate. Our focus is on accounts that delete a much larger volume.

Each day, we parsed the tweet deletion notices to identify accounts that deleted at least 10 tweets in the previous 24-hour period. 
For these accounts, we queried the Twitter user object API. 
A user object is returned if an account is active. Else, the API returns a notification with the reason it is unavailable (suspended or deleted). 
30 days after an account is deleted, Twitter deletes all of its tweets. This generates a large number of  deletions that are not interesting for the purposes of our analyses. Therefore, we discarded deleted accounts.
We queried the user object API rather than relying on account deletion notices in the CF because there may be a delay of more than 30 days between the time an account is deleted and the time when all of the account's tweets are deleted.
The user object also contains additional metadata, such as tweet counts and profile descriptions, that are analyzed in later sections. 

There was a misalignment of up to a few hours between the 24-hour interval in which CF notices were aggregated and the time when the user objects were queried from the API. This was due to two constraints. First, we had to wait until past midnight to determine which accounts deleted at least 10 tweets the prior day. Second, querying a large collection of user objects daily required throttling requests due to the Twitter API's rate limit. 

We analyzed deletion metadata for accounts that were still active or suspended within 24 hours of their tweet deletion notices. Suspended accounts are relevant to our analysis because they can delete tweets.  
Note that some accounts switched between being active and suspended. 
We use the terms \textit{active days} and \textit{suspended days} to characterize the periods in which accounts are in these states. 

Overall, our dataset includes metadata for almost 1.2B deletions by over 11.6M accounts (Table~\ref{tab:user_groups_classification}). 
We categorized accounts into categories based on whether they deleted tweets only occasionally (\textit{one-day deleters}) or daily (\textit{30-days deleters}) during the span of our collection. We also identified accounts that violated Twitter's limit of 2,400 tweets per day (\textit{suspicious deleters}), as explained later.

Additionally, we retrieved all the unlike notices during the data collection period for tweets included in our dataset.

\section{ESTIMATING THE NUMBER OF DELETED TWEETS}

Researchers who do not have access to the CF may wish to estimate the number of tweets deleted by an account in some time interval, for example to examine whether the account is engaging in suspicious activities.
This can be done by comparing the tweet counts at two times: if the count decreases, the difference can be used to estimate the number of deleted tweets~\cite{infodemic-twitter-vs-facebook,torres2020manufacture}.
For example, if an account has 500 tweets on Monday and 400 tweets on Tuesday, then we infer that it must have deleted at least 100 tweets between Monday and Tuesday. This estimate is a lower bound on the true number of deleted tweets --- the account might actually have deleted, say, 150 tweets and then posted 50 new ones, yielding the same difference of 100. If the number of tweets increases or stays constant, we cannot infer any number of deleted tweets; the account might have deleted nothing, or might have deleted fewer tweets than it posted during the time interval.

To assess the extent to which such estimates can accurately quantify deletion behaviors, let us compare the estimates of deleted tweets obtained by this approach with the actual numbers obtained from CF metadata. We start from the \textit{tweet count} (\texttt{statuses\_count}) field in the user object, which reports the total number of tweets $n^i(t)$ posted by an account $i$ at time $t$. 
Our comparison can leverage tweet counts and true deletion counts for each account and each day in our dataset.

Let us denote by $d^i_a(\Delta t)$ and $d^i_e(\Delta t)$ the actual and estimated \emph{daily} numbers of tweets deleted by account $i$ during time $\Delta t$, respectively (in our data, the maximum time resolution is $\Delta t = 1$ day). For the simple case in which tweet counts are available for two consecutive days $t$ and $t+1$, if $n^i(t+1) < n^i(t)$ and the true deletion count is above 10 for the interval $(t, t+1)$, we can use 
\begin{equation}
d^i_e(\Delta t) = n^i(t) - n^i(t+1).
\label{eq:conseq_days}
\end{equation}
Note that we only estimate the number of deleted tweets for accounts and days such that Eq.~\ref{eq:conseq_days} yields a positive estimate; as explained above, we cannot infer any deletions otherwise.

There are two more complicated ``gap'' cases such that $\Delta t = t_2 - t_1 > 1$: (1) we may not have actual deletion counts for one or more consecutive days, either because the account did not delete any tweets, or because they deleted fewer tweets than our threshold of 10; and/or (2) we may not have tweet counts for one or more consecutive days, because the account was suspended. In both of these cases we are still able to estimate the daily number of deleted tweets by considering the last day $t_1$ when a tweet count was available (the account was active), and the first day $t_2$ when the tweet count and actual deletion data are both available: 
\begin{equation}
d^i_e(\Delta t) = \frac{n^i(t_1) - n^i(t_2)}{t_2 - t_1}
\label{eq:gap_days}
\end{equation}
where $n^i(t_1) > n^i(t_2)$.

Note that while the tweet counts are obtained from the Twitter user object API, the accounts that we query are those that appear in the CF. Without access to the CF, one could still estimate the number of deletions from any collection of tweets. To explore such a scenario, we collected tweets from the 1\% streaming API for 24 hours between December 16--17, 2021. For each user $i$ with at least two tweets, we estimated the number of deleted tweets $d^i_e$ by comparing the tweet counts from the first and last tweet (Eq.~\ref{eq:conseq_days}). 

\begin{figure}[t]
    \centering
    \includegraphics[width=0.48\textwidth]{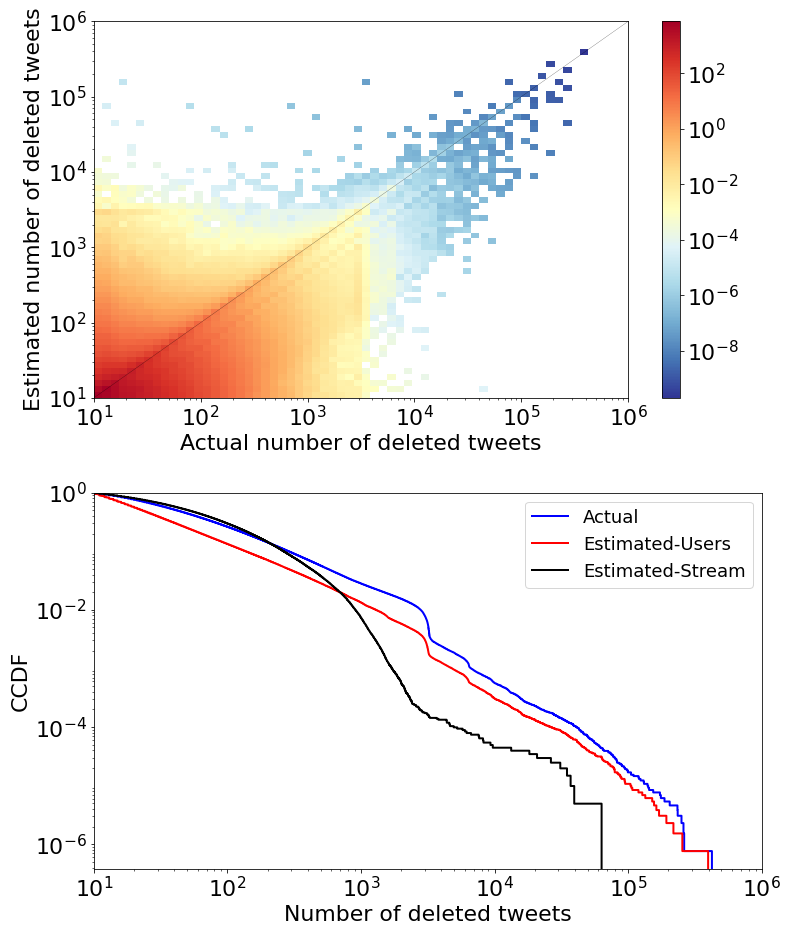}
    \caption{Comparison between estimated vs. actual numbers of daily deleted tweets. Top: Heat map of the joint distribution. Bottom: Complementary cumulative distribution functions (CCDF). The two estimated distributions are based on accounts in the CF (via user object queries) or in a random sample of tweets (via 1\% streaming API).}
    \label{fig:heatmap_interpolation}
\end{figure}

The estimates $d_e$ from Eqs.~\ref{eq:conseq_days} and \ref{eq:gap_days} and the actual daily numbers of deleted tweets $d_a$ are compared in Fig.~\ref{fig:heatmap_interpolation} (top) for all the  data points such that $d_e \ge 10$ --- approximately 1.3M  account/day pairs corresponding to over 748K accounts. We also aggregated the data by taking the median values for each user (not shown), yielding very similar results. 
The diagonal in the heat map  represents the cases where the estimates are exact. 
The higher density below the diagonal indicates that the  numbers of deleted tweets are typically underestimated. The presence of points above the diagonal (overestimates) is due to gap cases as well as the misalignment explained in the previous section. 

The cumulative distributions in Fig.~\ref{fig:heatmap_interpolation} (bottom) similarly demonstrate the underestimation. Here we also include the distribution of estimated deletions obtained from the random sample of tweets. Kolmogorov-Smirnov tests confirm that the distributions are significantly different ($p < 0.01$). The average number of deleted tweets is underestimated by 45\% when considering accounts in the CF (actual 171, estimated 94) and by 34\% when considering the random sample of tweets (estimated 113). However, the errors can be much larger, by orders of magnitude.
Similar results (not shown) are obtained by considering only cases of consecutive days (Eq.~\ref{eq:conseq_days}), i.e., ignoring gaps (Eq.~\ref{eq:gap_days}). 

In principle, one could obtain better measurements of deleted tweets without access to the CF. One could use the stream/filter API to retrieve all the tweets that a set of users are publishing in real time. Alternatively, user objects could be queried frequently to achieve better granularity in the difference between tweet counts. Both of these approaches require prior knowledge of which accounts should be monitored. Further, APIs limit the number of tweets that can be streamed in real time and the query rates for user objects. Consequently, neither of these approaches is feasible; therefore, research is hindered by lack of access to deletion data. 

\section{DELETION BEHAVIORS}

To focus on abuse involving deletions, we must understand what constitutes anomalous deletion behavior. We consider three signals: deletion volume, deletion frequency, and age of deleted content. 
These statistics can be extracted from the available data and provide interpretable signals about the behavior of deleting users. 
For example, accounts that delete recent tweets but not in high volume are not suspicious because they could be correcting mistakes or using the platform in an ephemeral way (such as Snapchat messages or Instagram stories). Similarly, accounts that delete old tweets in high volume but not often could be motivated by legitimate privacy reasons. 
On the other hand, deleting a large volume of tweets every day could signal platform abuse, especially when the deletions target new tweets. 
Therefore, we focus on accounts that delete in high volume or high frequency. 

\begin{figure}[t]
    \centering
    \includegraphics[width=0.48\textwidth]{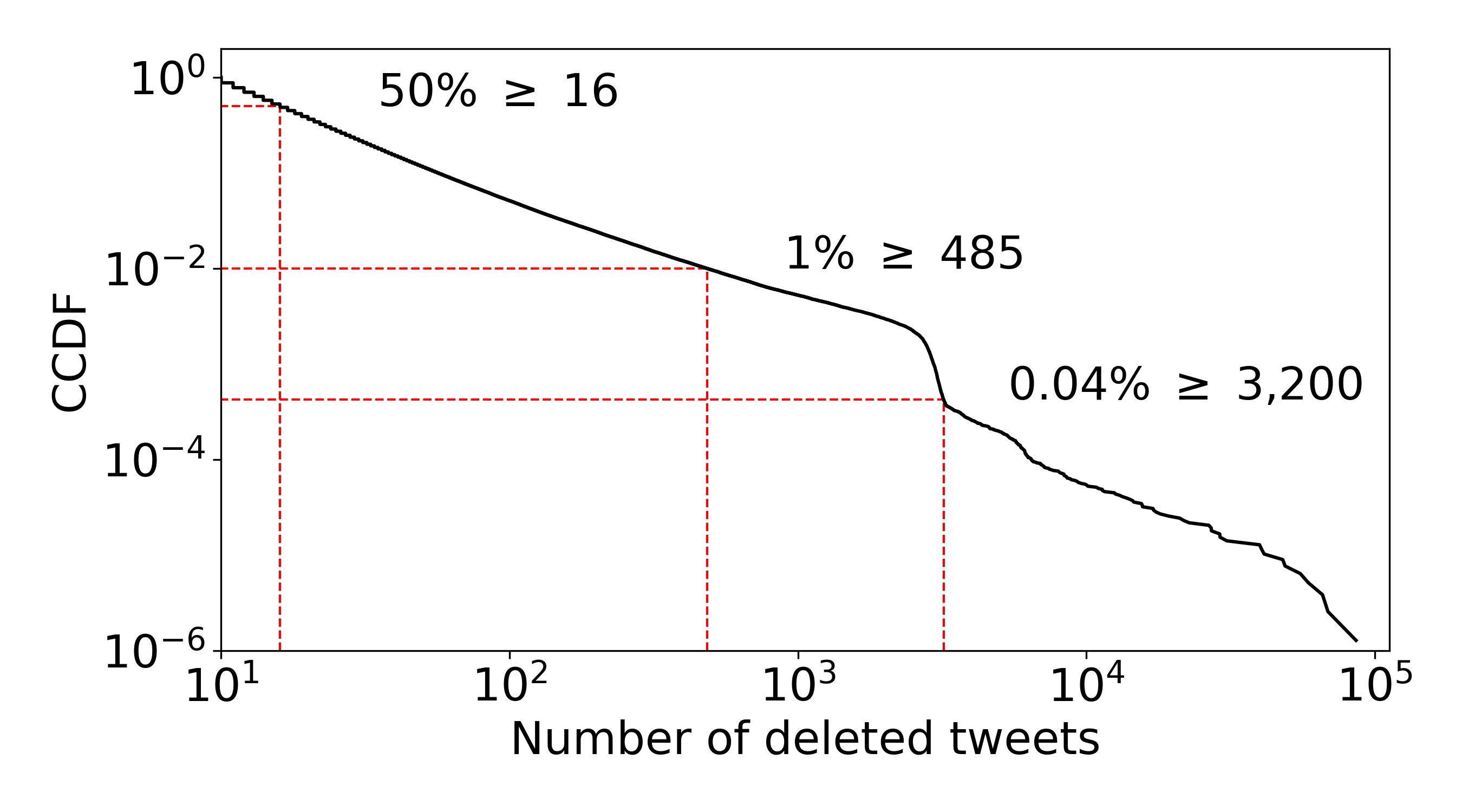}
    \caption{Complementary cumulative distribution of the number of deleted tweets per account that occurred on April 26, 2021. Other days produce similar distributions. 
    }
    \label{fig:ccdf_log_user_del}
\end{figure}

Each day, on average, over 914K accounts delete approximately 40M tweets. 
To characterize the typical deletion behaviors, let us consider the distribution of the number of tweets deleted daily by individual accounts.
Fig.~\ref{fig:ccdf_log_user_del} shows such a distribution for a single, typical day.  
(Recall that we only count deletions by accounts that delete at least 10 deletions in a day.)
The distribution has a heavy tail, with most accounts deleting few tweets (the median is 16 deleted tweets) and a small fraction of accounts deleting a very large volumes of tweets. For example, 0.04\% of accounts delete at least 3,200 tweets, which strongly suggests the use of deletion software. The rapid drop observable in the plot around 2,400--3,200 deletions is likely due to accounts reaching the maximum number of tweets that can be posted in a day (2,400) and deleting all of them, and accounts deleting the maximum number of tweets retrievable with the timeline API (3,200). Our results also show that these accounts tend to delete newer content.

\begin{figure}[t]
    \centering
    \includegraphics[width=0.48\textwidth]{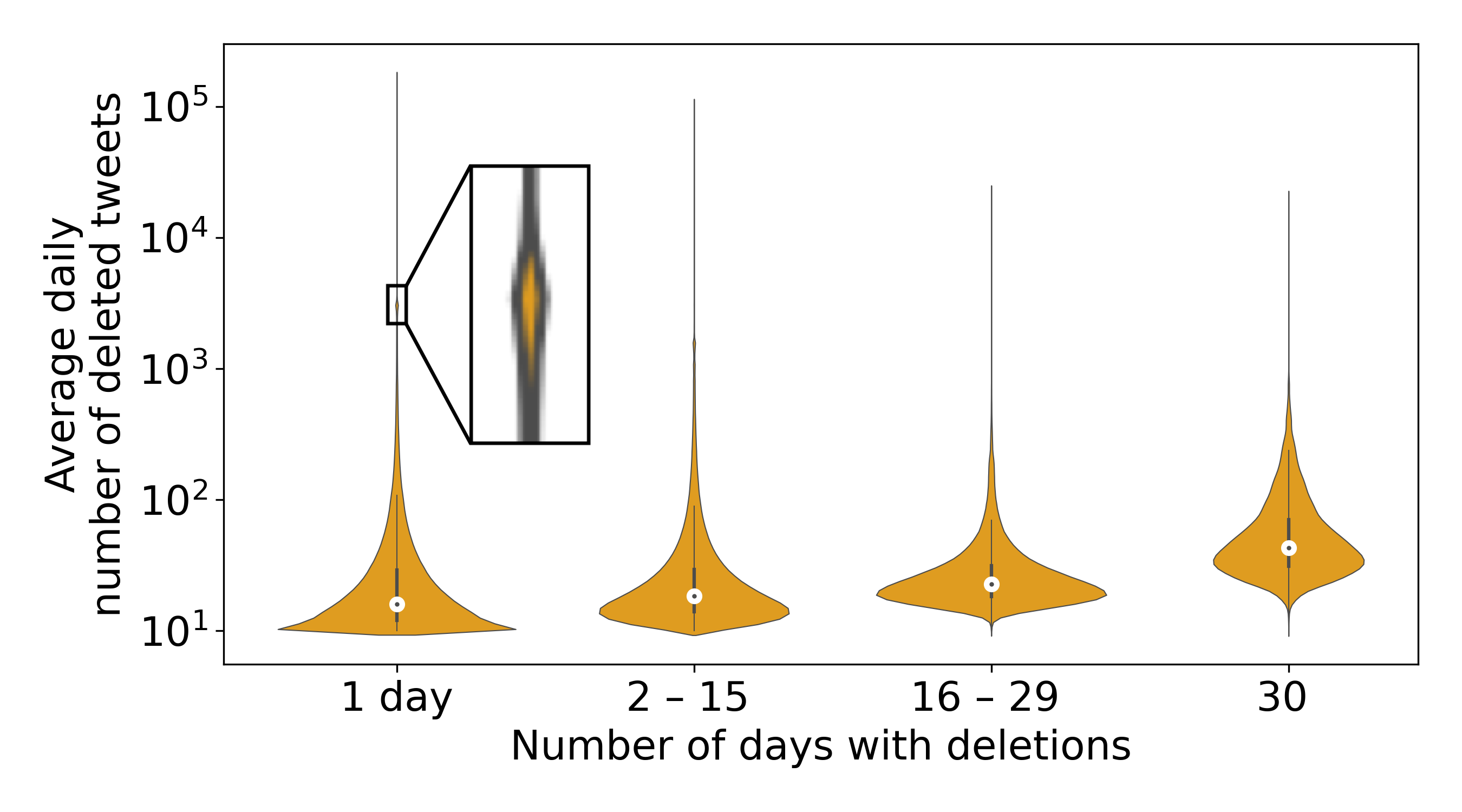}
    \caption{Distributions of the average daily numbers of deleted tweets per account, for groups of accounts that delete at different frequencies during our collection period. 
    Medians are shown as white circles.
    The highlighted region of the one-day distribution corresponds to Twitter's timeline API limit of 3,200 tweets.}
    \label{fig:total_days_deleting_tweets_violin}
\end{figure}

To provide additional context about  deletion behaviors, consider how the volume of deletions varies with the frequency at which the deletions occur.
Fig.~\ref{fig:total_days_deleting_tweets_violin} plots the distributions of average daily numbers of deleted tweets for accounts that delete at different frequencies (number of days). 
The highlighted peak in the one-day distribution corresponds to approximately 3,200 tweets --- the maximum number of tweets retrievable from Twitter's timeline API. We conjecture that mass-deletion tools use this API to obtain the IDs of the most recent tweets, creating a natural limit on the number of deletions.
Irrespective of deletion frequency, the medians of the distributions lie between 10 and 100 deleted tweets on average, but we again observe heavy tails with minorities of accounts responsible for majorities of tweet deletions in all groups. 
Overall, accounts that delete more frequently tend to delete more tweets: the median of the average daily number of deleted tweets is highest for \textit{30-days deleters}. 

\begin{figure}
    \centering
    \includegraphics[width=0.48\textwidth]{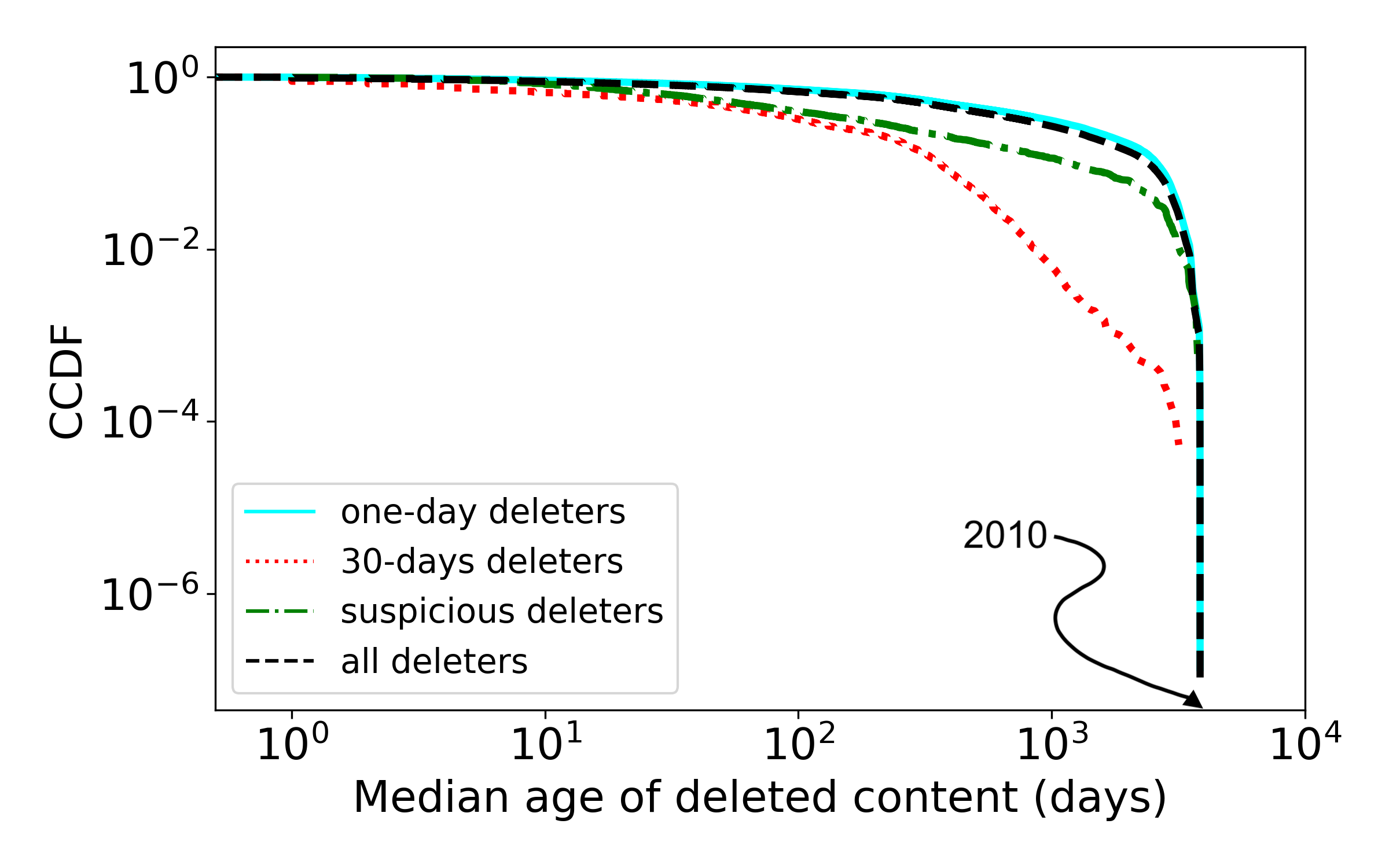}
    \caption{Complementary cumulative distributions of the median age of  content deleted by the four classes of accounts in Table~\ref{tab:user_groups_classification}. The oldest deleted tweets date back to 2010.}
    \label{fig:sup_median_delete_delta_days_tweets_age_density}
\end{figure}

Considering the age of deleted content, Fig.~\ref{fig:sup_median_delete_delta_days_tweets_age_density} shows that accounts that delete a lot tend to remove newer tweets. For example, further analysis shows that 50\% of \textit{one-day deleters} remove content with a median age of 375 days or more, compared to 39 days for \textit{30-days deleters} and 57 days for \textit{suspicious deleters}.

To summarize, we find evidence of accounts engaged in high-volume, high-frequency deletions of recent content. We believe these anomalous behaviors, which deviate from normal activity by orders of magnitude, are highly suspicious.

\section{ABUSIVE DELETION BEHAVIORS}

Let us explore two types of abuse identified from suspicious deletion behaviors: high-volume posting (flooding) and coordinated manipulation of content recommendation.

\subsection{Flooding}

One of the principal concerns of social media platforms is to balance a user's ability to contribute content while keeping the experience safe for others.
Twitter limits the daily number of tweets that can be posted by an account\footnote{\scriptsize\url{help.twitter.com/en/rules-and-policies/platform-manipulation}} as one of the measures aimed at safeguarding the user experience. 
In this regard, Twitter's documentation currently states\footnote{\scriptsize\url{help.twitter.com/en/rules-and-policies/twitter-limits}}:
\begin{quote}
\textit{``\textbf{Tweets}: 2,400 per day. The daily update limit is further broken down into smaller limits for semi-hourly intervals. Retweets are counted as Tweets.''}
\end{quote}
While this policy suggests that it is not possible to post more than 2,400 tweets per day, here we determine whether deletions can be exploited to circumvent this limit.

Suppose an account posts a tweet every six seconds, so that it would reach the limit after four hours. Now the account deletes 2,400 tweets, and then starts over. In this scenario, the account would circumvent the limit and post a total of 14,400 tweets in a day. (This is a simplified scenario, as Twitter has more undisclosed limits for shorter durations.)

To check if this kind of abuse exists, we can estimate the total number of tweets posted by account $i$ in a day by adding two numbers: the  difference in tweet counts $N^i(t)=n^i(t) - n^i(t-1)$, obtained from user objects across consecutive days, and the number of deleted tweets in the corresponding time interval $d^i_a(t-1,t)$, obtained from the CF. 

Recall that $N^i(t)$ can be negative, indicating that the number of deleted tweets is larger than the number of posted tweets in the same day. If instead $N^i(t) > 0$, then the account posted more tweets than it deleted. 
In either case, and irrespective of when the deleted tweets were originally posted, $N^i(t) + d^i_a(t-1,t)$ provides the total number of tweets posted during day $t$. 
As an illustration, say  account $i$ deleted 100 tweets and posted 80 tweets during day $t$. Then $N^i(t) = -20$ and $d^i_a(t-1,t) = 100$. The sum gives the actual number of posted tweets, i.e., 80.

To perform this analysis, we identified accounts whose tweet counts could be accessed on consecutive days from user objects in our dataset. Note that the tweet count measurements are delayed with respect to the deletion counts, due to the misalignment discussed earlier. This means that we have approximate values of  the tweet counts and therefore of the total number of tweets posted.  

\begin{figure}[t]
    \centering
    \includegraphics[width=0.48\textwidth]{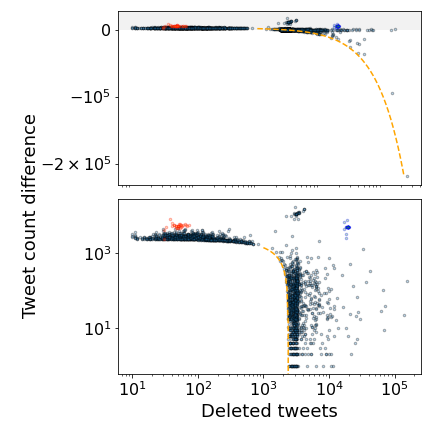}
    \caption{Tweet count difference $N$ vs. number of deleted tweets $d_a$ for \textit{suspicious deleters} that exceeded the limit of 2,400 tweets posted in a single day. The dashed line represents the threshold $N + d_a = 2{,}400$ tweets. Top: log-linear plot showing all points corresponding to account/day pairs circumventing the limit. Bottom: region with positive $N$ (highlighted in the top plot) using a log-log scale. Red points correspond to \texttt{@AmazonHelp} (see Discussion section), while blue points correspond to an account that publishes coded content (see Suspensions section for details). }
    \label{fig:2400tweets}
\end{figure}

We identified 1,715 \textit{suspicious deleters} that violated the 2,400-tweets limit at least on a single day, and 120 accounts that do so more than once.
Fig.~\ref{fig:2400tweets} plots these violations, each corresponding to an account/day pair ($i$, $t$).
The x-axis represents the number of deleted tweets $d_a^i(t)$ and the y-axis represents the difference in tweet counts $N^i(t)$. 
All violation points are above the line $N + d_a = 2{,}400$ tweets. 

\begin{figure}
    \centering
    \includegraphics[width=0.48\textwidth]{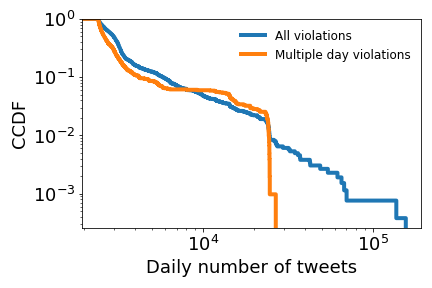}
    \caption{Complementary cumulative distributions of the numbers of daily tweets exceeding the 2,400-tweet limit.}
    \label{fig:2400tweetsdist}
\end{figure}

In Fig.~\ref{fig:2400tweetsdist} we visualize the distribution of the number of daily tweets $N^i + d_a^i$ for \textit{suspicious deleters} $i$ that exceed the daily limit. 
We observe cases of an account posting hundreds of thousands of new tweets in a day. From Fig.~\ref{fig:2400tweets} we can tell that these are cases in which the bulk of evidence originates from large numbers of deleted tweets, which are not matched by decreases in tweet counts.
Based on our observations, we suspect these may be anomalies caused by  stale tweet count data provided by the Twitter API when information about deleted tweets has not yet propagated to the queried data center. 
To focus on the most likely violators, Fig.~\ref{fig:2400tweetsdist} also shows the distribution of the number of daily tweets for the subset of \textit{suspicious deleters} that violate the limit multiple times. Although the tail of the distribution is truncated, we still observe evidence of multiple accounts violating Twitter's daily limit, sometimes by an order of magnitude (posting over 26K tweets).

\subsection{Coordinated manipulation}

The \textit{like} button is a way for Twitter users to signal a positive sentiment toward a tweet.
Such signals are used by the platform's feed ranking and recommendation algorithm: tweets and users with many likes are prioritized.\footnote{\scriptsize\url{blog.twitter.com/engineering/en_us/topics/insights/2017/using-deep-learning-at-scale-in-twitters-timelines}}
Therefore, likes are a potential vector of attack for malicious accounts seeking to amplify content and increase influence by manipulating the platform.

The capacity to study likes is severely constrained by the rate limits imposed by the Twitter API.
However, the deletion stream emits a compliance notice when a like is removed from a tweet. This occurs when an account \textit{unlikes} a tweet or the liked tweet is deleted. 
There are cases in which an unlike is warranted, such as when a user retracts an accidental like.
However, it is not generally expected behavior for an account to repeatedly like and unlike a tweet in short order or over long periods of time.
Both of these could be interpreted as signals of inauthentic behavior by an account.

\begin{figure}
    \centering
    \includegraphics[width=0.45\textwidth]{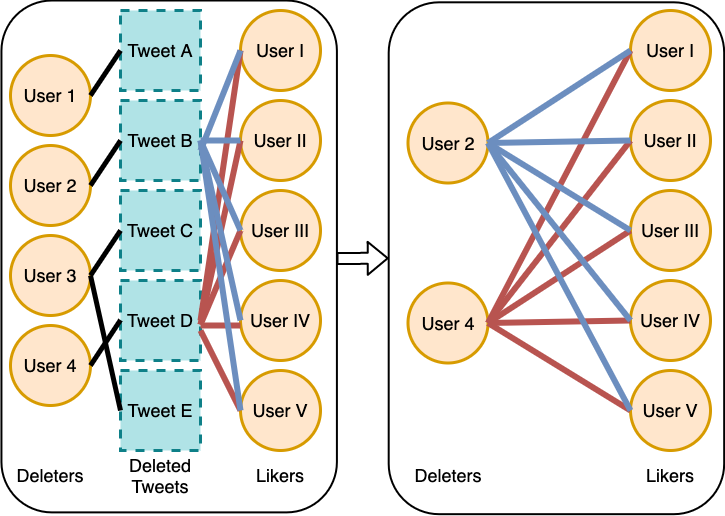}
    \caption{This diagram illustrates how we connect interactions between accounts using CF notices. Left panel: We construct a tripartite network consisting of deleted tweets (center), accounts that deleted those tweets (left), and accounts that liked those tweets (right). Right panel: By projecting the network onto deleters and likers, we get a bipartite network that connects interacting accounts.}
    \label{fig:coordination_explanation}
\end{figure}

To identify groups of accounts likely involved in coordinated liking and deleting behaviors, we selected accounts with multiple unlikes for the same tweet that is eventually deleted. We adopted a methodology similar to~\citet{Pacheco2021Coordinated},  modified to start from a tripartite network as shown in Fig.~\ref{fig:coordination_explanation}. 
This network consists of \textit{deleters}, \textit{deleted tweets}, and \textit{likers}.
The edges between deleted tweets and likers have weights that account for the number of unlikes.
To focus on the most unusual unlikers, we removed 93.5\% of accounts that had unliked the same tweet less than five times --- assuming that the like icon could be tapped a few times accidentally. 
We then projected the tripartite network into an unweighted, directed, bipartite network connecting likers to deleters.
Finally, we filtered out weakly connected components with less than 10 nodes to focus on the 1.7\% most suspicious clusters. These represent coordination scenarios in which multiple accounts repeatedly like, unlike, and eventually delete one or more tweets. 

\begin{figure*}[t]
    \centering
    \includegraphics[width=0.9\textwidth]{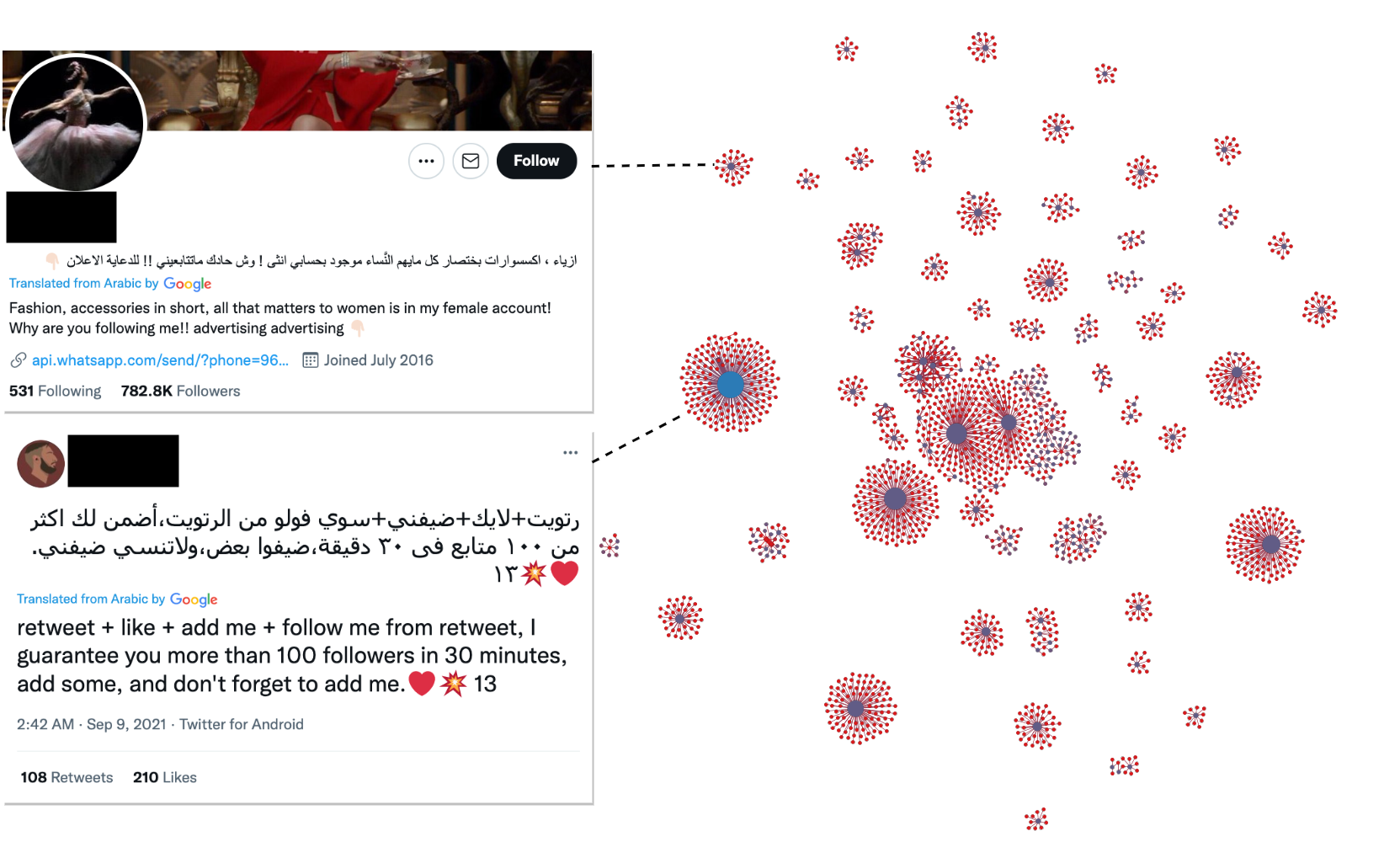}
    \caption{Coordinated network of accounts that repeatedly liked and unliked tweets that were eventually deleted. Node size is proportional to in-degree. Blue and red nodes represent accounts that are mostly deleters and mostly likers, respectively (see text). The screenshots highlight the center nodes of two of the networks.}
    \label{fig:coordination}
\end{figure*}

The resulting coordinated networks can be seen in Fig.~\ref{fig:coordination}. 
Since an account can be a liker as well as a deleter, we calculated the ratio between the number of unlikes and deletions for each node, and used this ratio to color the nodes in the figure. Most coordinated networks have a deleter (blue hub) and several likers (red spokes).

For illustration purposes, we highlight two clusters in Fig.~\ref{fig:coordination} with screenshots of a hub profile and a tweet, respectively. These were obtained from the Twitter website and not from deleted content.
The top account self-identifies as fashion/advertisement-related with hundred of thousands of followers.
The bottom tweet promises users more than 100 followers in 30 minutes if they complete some tasks.
These accounts seem to be engaging in inauthentic behavior that likely violates platform rules.

\section{CHARACTERIZATION OF DELETERS}

In this section we analyze the profiles extracted from the Twitter API to characterize some of the  accounts that delete frequently and the suspicious accounts that circumvent the Twitter limit, in an effort to gain additional understanding of how these malicious actors operate. 

\subsection{Profiles}

\begin{figure}[t]
    \centering
    \includegraphics[width=.495\columnwidth]{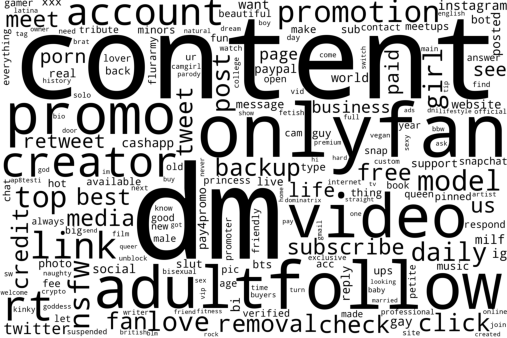}
    \includegraphics[width=0.495\columnwidth]{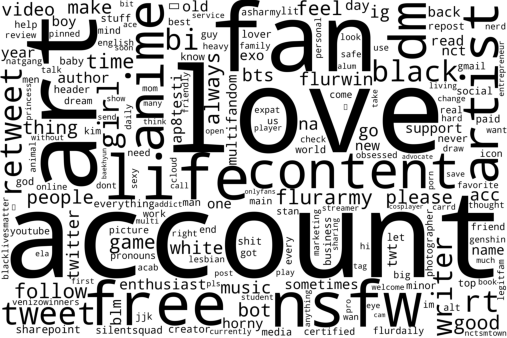}\\
    \vspace{1mm}
    \includegraphics[width=\columnwidth]{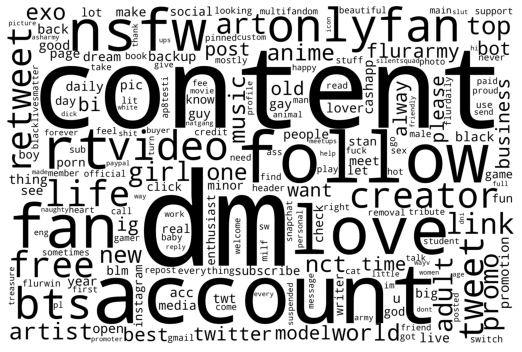}
    \caption{Word clouds of the text contained in profile descriptions for the top 10\% of \textit{30-days deleters} (top left), bottom 10\% of \textit{30-days deleters} (top right), and \textit{suspicious deleters} (bottom).}
    \label{fig:wordcloud}
\end{figure}

Account profiles often include information about the owner and/or their interests. The top two languages of the profile descriptions of the \textit{suspicious deleters} and \textit{30-days deleters} are English and Japanese. Let us focus on English profiles (those in Japanese have similar content). Fig.~\ref{fig:wordcloud} illustrates common terms from the profiles of the top/bottom 10\% of \textit{30-days deleters} as well as \textit{suspicious deleters}. 
Some of these terms (e.g., \textit{follow}, \textit{backup}, \textit{promo}) could be associated with activities that are restricted on the platform.\footnote{\scriptsize\url{help.twitter.com/en/rules-and-policies/platform-manipulation}}
The top 10\% of \textit{30-days deleters} and the \textit{suspicious deleters} appear to exploit deletions to produce such problematic content without being detected, warranting further scrutiny.

\subsection{Automation}

\begin{figure}[t]
    \centering
    \includegraphics[width=0.45\textwidth]{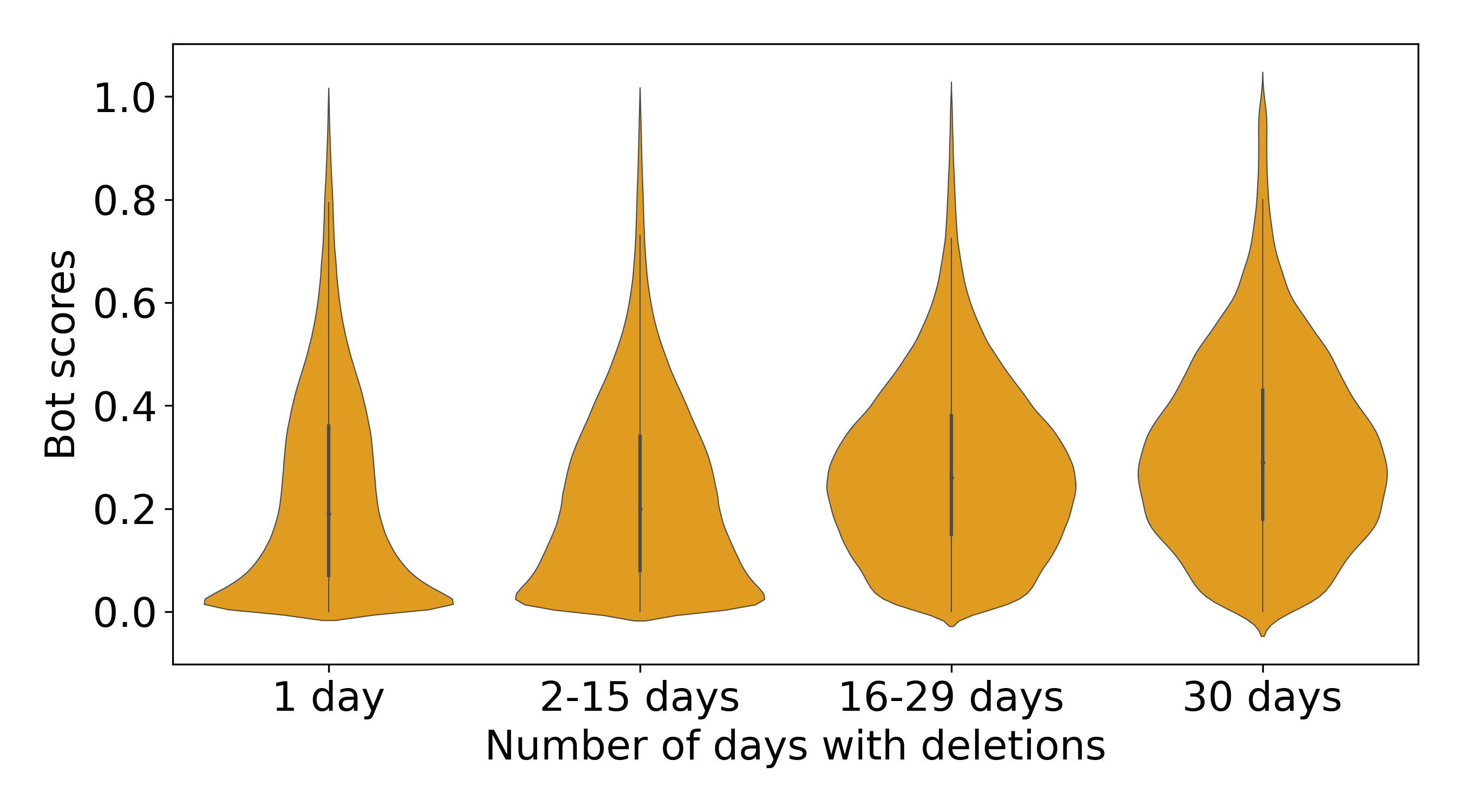}
    \caption{Distributions of bot scores for groups of accounts that deleted tweets  between 1 and 30 days.}
    \label{fig:botscores}
\end{figure}

Let us use the \textit{BotometerLite} tool to estimate the prevalence of bot-like accounts in our dataset. BotometerLite is a machine learning model that was trained using datasets of bot- and human-operated accounts. The model has been shown to achieve high accuracy in cross-validation and cross-domain evaluations~\cite{yang2020scalable}. 
For each account, BotometerLite generates a score between zero and one, with zero indicating a more human-like account and one indicating a more bot-like account. 
We grouped the accounts as in Fig.~\ref{fig:total_days_deleting_tweets_violin}, based on the number of days in which each account deleted tweets. 
Fig.~\ref{fig:botscores} shows the distributions of bot scores for accounts in each group. 
We observe that accounts that delete more frequently tend to have higher bot scores. 

\subsection{Suspensions}

For \textit{suspicious deleters} and accounts involved in coordinated manipulation, we queried their user objects on September 12, 2021 using Twitter's API V2 to see if Twitter had taken any enforcement action against them. We also queried \textit{one-day deleters} and \textit{30-days deleters} as points of reference. The suspension statistics are shown in Table~\ref{table:suspensions}. Very few of the \textit{suspicious deleters} have been suspended, suggesting that Twitter has not clamped down on this kind of abuse. 

\begin{table}[t]
\small
\centering
\caption{Suspension statistics.}
\begin{tabular}{lrrr} 
 \hline
 \textbf{Account type} & \textbf{Suspended} & \textbf{Total} & \textbf{Percentage} \\
 \hline
 Coordinated & 379 & 1,884 & 20.1\% \\
 One-day deleters & 579,933 & 7,938,077 & 7.3\% \\
 30-days deleters & 1,725 & 27,065 & 6.4\% \\ 
 Suspicious deleters & 13 & 1,715 & 0.76\% \\ \hline
\end{tabular}
\label{table:suspensions}
\end{table}

\begin{figure}[t]
    \centering
    \includegraphics[width=0.8\columnwidth]{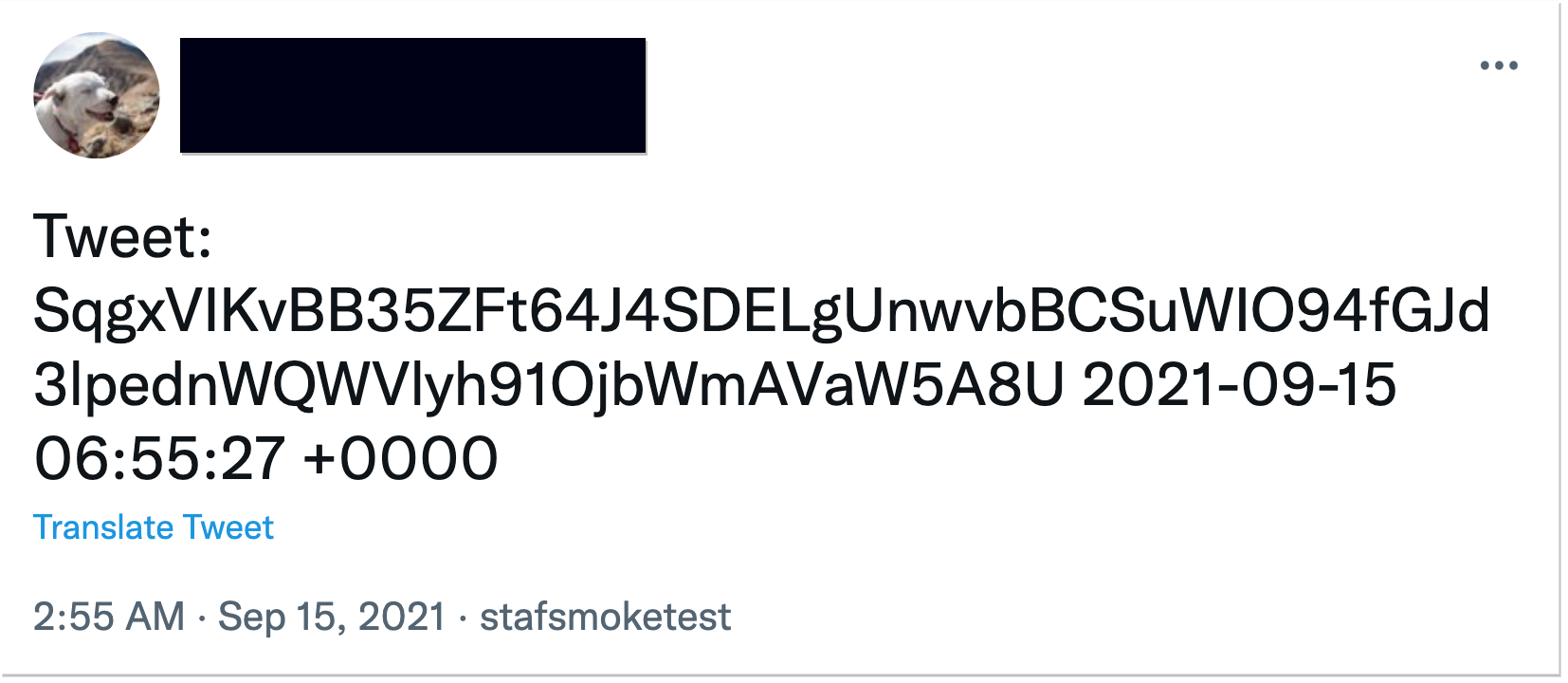}
    \caption{Screenshot of a typical tweet by a suspicious account that tweets and deletes a high volume of coded content.}
    \label{fig:mrgiggles}
\end{figure}

Among the accounts that are not suspended as of this writing and that exhibit high volumes of posts and deletions, we identified two that tweet coded content. A tweet by one of such accounts is shown in Fig.~\ref{fig:mrgiggles}. This account (also highlighted in Fig.~\ref{fig:2400tweets}) was created in August 2018 and, despite deleting large numbers of tweets, still has 7.5 millions tweets that have not been deleted to date.

\section{DISCUSSION}

We presented the first exhaustive, in-depth study of the deletion behaviors of Twitter users to identify cases of platform abuse. We started our analysis by assessing the degree to which we could estimate the number of deleted tweets with tweet counts, which are publicly available, given that the CF service is not freely available. 
Our findings illustrate that the use of tweet counts to estimate the number of deleted tweets results in an undercount of the actual number of deletions.

We focused on abusive deletion behaviors showing that while most accounts (68\% of all users) deleted tweets only on a single day, a small subset (0.23\%) deleted tweets for all 30 days of our collection period. A smaller subset (0.04\%) deleted on average at least 3,200 tweets daily.

We also identified two cases of platform abuse on Twitter. 
First, we identified 1,715 users that utilize deletions to post over 2,400 tweets per day, which exceeds Twitter's daily limit. As shown in Table~\ref{tab:user_groups_classification}, these suspicious accounts represent a tiny fraction of deleters (0.015\%), but delete a disproportionately large number of tweets (over 1\%). 
As a caveat, we should note that some accounts might have a special arrangement with Twitter allowing them to post more than 2,400 tweets per day. For example \texttt{@AmazonHelp} is one of these (see Fig.~\ref{fig:2400tweets}). 

Second, we uncovered inauthentic behaviors consisting of accounts that coordinate to repeatedly like and unlike a tweet before it is eventually deleted.
We suspect that further study of all unlikes (not just unlikes of deleted tweets) could lead to the identification of a significantly higher number of abusive accounts. This could provide visibility into the operations of commercial metric inflation services.

Our study is limited by some technical aspects of the data collection process, as well as noise in data provided by the Twitter API. Both of these issues can affect the number of tweets, and therefore add noise to our downstream analyses to estimate the numbers of deletions and posts.  

Despite these limitations, we have shown that content deletion metadata can provide valuable insights into a thus-far neglected form of platform abuse. 
Access to deleted content would make it possible to study the impact of malicious deletion behavior. One could also explore whether content generated by suspicious deleters contains patterns that could be used to distinguish them from other users. However, such access is prohibited by platform terms of service in order to protect user privacy. This creates a conflict between the need to protect user privacy and the need to understand and combat platform abuse.

Twitter now makes compliance data available for queried tweets and users.\footnote{\scriptsize\url{developer.twitter.com/en/docs/twitter-api/compliance/batch-compliance/introduction}} We believe that platforms should make deletion metadata streams and ---with proper privacy safeguards--- deleted content available to academic researchers to enable further investigations that could lead to a safer experience for users~\cite{misinfo_data}.

\subsection{Ethics statement} 

The data analyzed here is provided by Twitter under a license that prohibits redistribution. For this reason, as well as to honor user intent, we are not allowed to make it publicly available. 
Also to respect user privacy, we only examine deletion metadata and not deleted content. 
This ensures that our use of the compliance data is consistent with Twitter guidelines. 
Our analysis of public Twitter content is exempt from IRB review (Indiana University protocol 1102004860).

\subsection{Acknowledgments}

The authors are grateful to Sadamori Kojaku for his help in translating Japanese terms found in the word cloud.
This work was supported in part by Knight Foundation, Craig Newmark Philanthropies, and DARPA (contracts W911NF-17-C-0094 and HR001121C0169).
Any opinions, findings, and conclusions or recommendations expressed in this material are those of the authors and do not necessarily reflect the views of the funding agencies.

\bibliographystyle{aaai}
\bibliography{refs.bib}
\end{document}